\documentclass[11pt]{article}

\usepackage{sectsty}
\usepackage{graphicx}
\usepackage{cite}
\usepackage{url}
\usepackage{lscape}
\usepackage{indentfirst}
\usepackage{latexsym}
\usepackage{multirow}
\usepackage{tabls}
\usepackage{wrapfig}
\usepackage{longtable}
\usepackage{supertabular}
\usepackage{caption}
\usepackage{subcaption}
\usepackage{amsmath}
\usepackage{amssymb}
\usepackage{setspace}
\usepackage[bookmarks, colorlinks=true, plainpages=false, citecolor=blue, urlcolor=blue, filecolor=blue, linkcolor=blue]{hyperref} 
\usepackage[font={footnotesize,stretch=1.0}]{caption}

\allsectionsfont{\centering}
\newcommand{\pd}[2]{\frac{\partial #1}{\partial #2}}

\newcommand{\dv}{\, \mathrm{d}^3 \mathbf{v}}

\newcommand{\dom}{\, \mathrm{d}^2 \boldsymbol{\Omega}}

\newcommand{\vref}{v_\mathrm{ref}}

\newcommand{\lb}{\textsc{LightningBoltz}}
\newcommand{\urelvec}{\mathbf{u}_\mathrm{rel}}
\newcommand{\urel}{u_\mathrm{rel}}
\newcommand{\ucm}{\mathbf{u}_\mathrm{CM}}

\usepackage[margin=1.0in]{geometry}
\usepackage{authblk}
\usepackage{lineno}

\title{Multidisciplinary benchmarks of a conservative spectral solver for the nonlinear Boltzmann equation}

\author[1]{George J. Wilkie}
\author[2]{Torsten Ke{\ss}ler}
\author[2]{Sergej Rjasanow}
\affil[1]{Princeton Plasma Physics Laboratory, Princeton, NJ, USA}
\affil[2]{Saarland University, Saarbr\"ucken, Germany}

\begin{document}

\maketitle

\begin{abstract}
   The Boltzmann equation describes the evolution of the phase-space probability distribution of classical particles under binary collisions. Approximations to it underlie the basis for several scholarly fields, including aerodynamics and plasma physics. While these approximations are appropriate in their respective domains, they can be violated in niche but diverse applications which require direct numerical solution of the original nonlinear Boltzmann equation. An expanded implementation of the Galerkin--Petrov conservative spectral algorithm is employed to study a wide variety of physical problems. Enabled by distributed precomputation, solutions of the spatially homogeneous Boltzmann equation can be achieved in seconds on modern personal hardware, while spatially-inhomogeneous problems are solvable in minutes. Several benchmarks against both analytic theoretical predictions and comparisons to other Boltzmann solvers are presented in the context of several domains including weakly ionized plasma, gaseous fluids, and atomic-plasma interaction.
\end{abstract}

\section{Introduction}

The statistics of many mutually-interacting classical particles are formally reduced to the Boltzmann transport equation under the assumptions of molecular chaos and binary collisions which occur instantaneously. It is a nonlocal, nonlinear, high-dimensional partial integro-differential equation for the scalar distribution function which is notoriously difficult to solve directly. Over the past 150 years, approximations to the Boltzmann equation have formed the basis of several scholarly fields including plasma physics and aerodynamics, and has broad direct application in many other fields including astrophysics, solid state physics, and semi-classical quantum field theory. However, there remain important problems in these fields which violate these assumptions, and only recently has direct solution of the nonlinear Boltzmann equation in non-ideal conditions become computationally feasible. We demonstrate that the conservative spectral algorithm of Gamba, Rjasanow~\cite{gamba_galerkinpetrov_2018} and Ke{\ss}ler~\cite{kesler_fully_2019}, when appropriately generalized, robustly reproduces known analytic and computational results in a wide variety of physical applications. The resolution required to pass these benchmarks is relatively low, but more demanding applications are enabled by precomputation and online storage of the discrete collision operators. Therefore, problems in the transitional Knudsen regime that have been hitherto been avoided due to the immense computational cost are now accessible on modest modern hardware. 

The Boltzmann transport equation solves for $f^{(s)} \left(\mathbf{r},\mathbf{v},t \right)$, the scalar probability distribution to find a particle of species $s$ at a location within $\mathbf{r} +\mathrm{d}\mathbf{r}$ at a velocity within $\mathbf{v} + \mathrm{d}\mathbf{v}$ at time $t$. The equation forms the basis of kinetic theory and is difficult to solve outside of idealized conditions. Further reduction is profitable for the cases of: long-range small-angle Coulomb interaction between charged particles or dominance of collisions\cite{landau_kinetic_1936,braginskii_transport_1965,chapman_s._c._mathematical_1990}. These approximations have been used to solve many important physical problems, but inevitably, problems arise that are outside the applicable domain of these approximations. Such applications include plasmas and gases which are marginally coupled, ionized, and/or collisional. In addition, the Boltzmann equation or its variants appear in several other domains of physics which are not fundamentally concerned with classical particle interactions, such as solid state physics\cite{ziman_electrons_2001}, and the semi-classical quantum field theory\cite{drewes_boltzmann_2013}.

Direct-simulation Monte Carlo (DSMC) has been a flagship technique for solving the Boltzmann equation\cite{bird_molecular_1994,nanbu_direct_1980}. This mature and robust stochastic method continues to benefit from advances in analysis and techniques\cite{galitzine_accuracy_2014,bird_forty_2001} and is especially well-suited to the free molecular flow regime. Nonlinear collisions are handled by sampling the evolving velocity distribution. The main source of error in DSMC is discrete sample noise, which becomes especially problematic at low Knudsen number. In the context of fusion particle exhaust, it was found that deterministic methods are significantly more efficient at equivalent simulation error with simplified collision operators\cite{tantos_deterministic_2020}. Furthermore, when the stochastic calculation is coupled to a deterministic solution, otherwise tolerable stochastic error significantly interferes with convergence to a self-consistent solution\cite{joseph_coupling_2017}. A direct comparison between DSMC and the conservative spectral method was presented in Ref.~\cite{kesler_fully_2019}.

While multi-dimensional phase-space advection alone is a nontrivial problem for deterministic methods\cite{juno_discontinuous_2018}, the primary difficulty in solving the Boltzmann equation is the collision operator, which takes the form of a bilinear, nonlocal, five-dimensional velocity integral. In general, if velocity space is discretized with $N$ degrees of freedom, then the discrete collision operator has $N^3$ elements, each of which is at least a five-dimensional integral. Therefore, in assessing methods of solving the nonlinear Boltzmann equation, the reduction of $N$ is of primary concern. Spectral methods therefore arise as a natural choice to minimize the degrees of freedom to represent velocity space, and significant progress in applied mathematics has identified candidate methods. For several decades, spherical harmonics have been widely recognized as a useful representation of velocity angles, and this forms the basis for solving the linear Boltzmann equation with so-called two-term\cite{pitchford_extended_1981} methods (and their multi-term extensions\cite{yachi_multi-term_1988}), which are widely used for low temperature plasma applications. Fourier representations of velocity space are also common and useful\cite{pareschi_fourier_1996,filbet_solving_2006}. With a finite spectral basis, inaccuracies are inevitable when the goal is the minimize $N$. When increasing $N$ arbitrarily is not an option, one is then left a choice: which properties of the distribution or collision operator to conserve discretely: phase-space volume, collisional invariants, positivity, and/or dissipation of entropy\cite{filbet_analysis_2011}?

\begin{figure}
    \centering
    \includegraphics[width=0.5\textwidth]{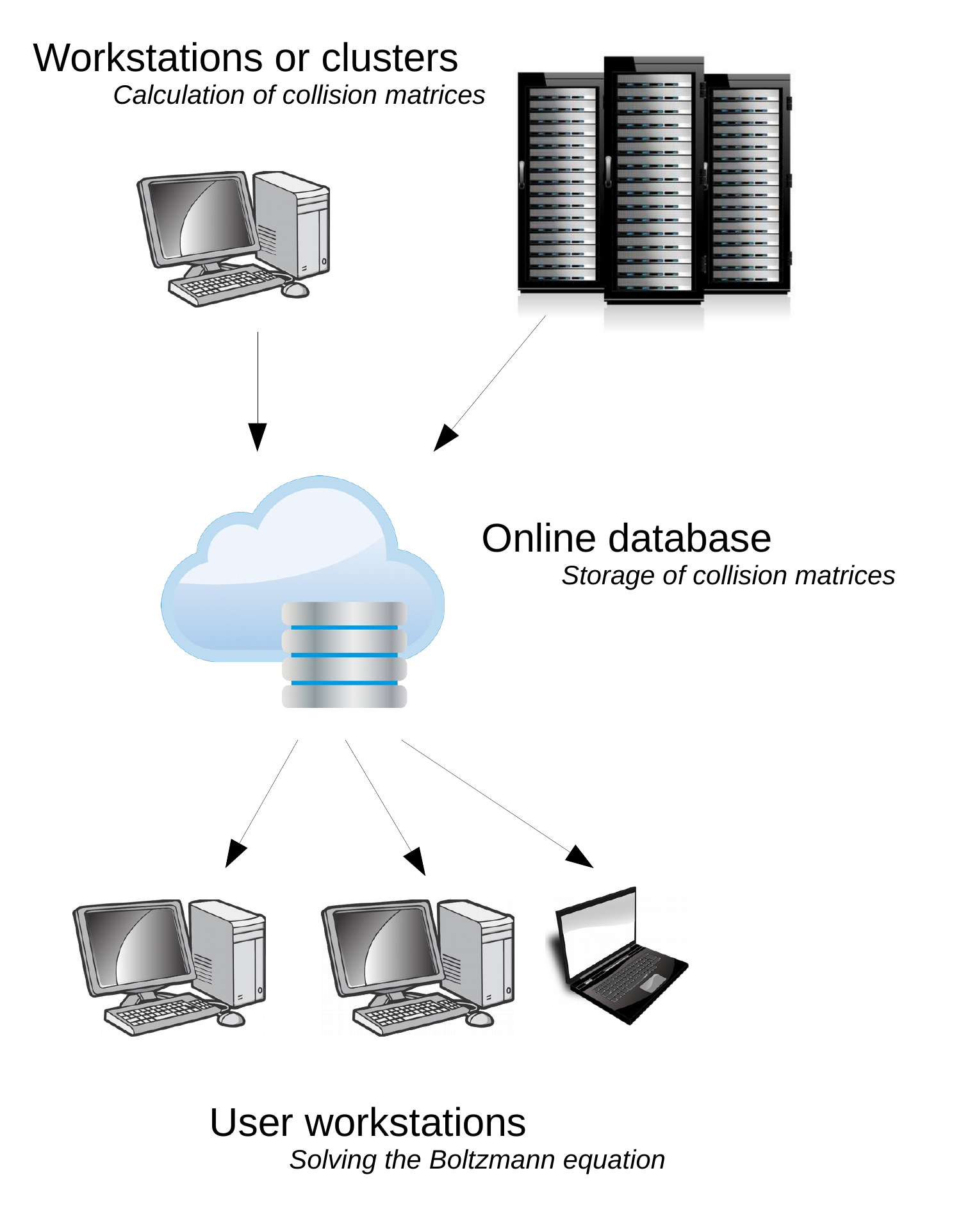}
    \caption{Schematic of the structure of {\lb}.}
    \label{architecture}
\end{figure}

This work adopts the conservative Galerkin--Petrov method put forth by Gamba, Rjasanow, and Ke{\ss}ler\cite{gamba_galerkinpetrov_2018,kesler_fully_2019}. In this scheme, the distribution function is expanded in a Burnett basis\cite{burnett_distribution_1936}, with test functions chosen to manifestly conserve collisional invariants. The Julia\cite{bezanson_julia_2017} implementation discussed here is referred to as {\lb}: a parallelized solver that expands the Galerkin--Petrov method for general/tabulated cross sections, inelastic collisions, force-field acceleration, sources, sinks, expanded integration techniques, and implicit time-stepping. When computed, the discrete collision operator is stored in an online database, which can be called upon for efficient direct solution of the Boltzmann equation. Figure \ref{architecture} shows a schematic of the numerical architecture. Discrete collision operators are calculated either \emph{en masse} on parallel clusters or on demand on local workstations. These are stored online and are downloaded if available when the user runs {\lb}.

In the following section, the implemented numerical technique is presented. In section \ref{benchmarksec}, several benchmarks are shown against both theoretically idealized cases and independent software. Section \ref{discsec} discusses some of the limitations of this method and how they can be overcome to provide a framework for a general domain-agnostic nonlinear Boltzmann solver.

The discrete form of the collision operator in the spectral scheme is a rank-3 tensor with $N^3$ elements. Each of these is an 8-dimensional integral: two for scattering solid-angle, and three each for integrations over $\mathbf{v}$ and $\mathbf{v_*}$.
Once calculated, the elements of the discrete collision operator are stored in an online database.

\section{Algorithm Description} \label{methodssec}

 In its present form, {\lb} solves the Boltzmann equation in three-dimensional velocity space and one dimensional configuration space.
Under this restriction, the Boltzmann equation is expressed as:
\begin{equation} \label{BE}
   \pd{f^{(s)}}{t} + v_z \pd{f^{(s)}}{z} + \frac{F_z}{m} \pd{f^{(s)}}{v_z} = S - L f^{(s)} + \sum\limits_{s_*} C_{ss_*},
\end{equation}
where $F_z$ is a force, $L$ is a loss term, $S=S\left(\mathbf{v},z \right)$ is a particle source, and $C_{ss_*}$ is the collision operator between species $s$ and $s_*$. The sum is over all species $s_*$ that species $s$ may interact with, including $s$ itself. Collisions are described by a change in the relative velocity $\urelvec$ in the center of mass frame moving at $\ucm = \left(m \mathbf{v} + m_* \mathbf{v}_* \right)/\left(m+m_*\right)$ such that the before the collision, the particles have velocities $\mathbf{v}' = \ucm + \left(\mu/m \right) \urelvec'$ and $\mathbf{v}_*' = \ucm - \left(\mu/m_*\right) \urelvec'$, with reduced mass $\mu = mm_*/\left(m+m_*\right)$. The velocities before or after\footnote{The primed velocities represent the pre-collisional velocities in the first (gain) term of the Boltzmann collision operator. In the second (loss) term, they represent the post-collision velocities. This has a tendency to cause confusion, especially when symmetries are used to swap primed and unprimed velocities and when expressing the weak form compactly. See Ref. \cite{hu_fast_2019} for a detailed exposition.} the collision are:
\begin{align}\label{vprime}
   \mathbf{v}' &= \ucm + \urel' \boldsymbol{\Omega} \\
   \mathbf{v}'_* &= \ucm - \urel' \boldsymbol{\Omega}
\end{align}
where $\urelvec'$ is the pre-collision relative velocity (for elastic collisions, $\urel'=\urel$) and $\boldsymbol{\Omega}$ is a unit vector. The scattering angle is $\chi = \mathrm{sin}^{-1}| \boldsymbol{\Omega} \times \urelvec / \urel |$.

The collision operator for nonlinear elastic scattering is:
\begin{equation} \label{elasticcollop}
   C_{(\mathrm{el})}\left[f^{(s)}, f^{(s_*)} \right] = \int \int \urel \sigma_{s,s_*}(\urel,\chi) \left[ f^{(s)}\left(\mathbf{v}'\right) f^{(s_*)}\left(\mathbf{v}'_*\right) - f^{(s)}\left(\mathbf{v}\right) f^{(s_*)}\left(\mathbf{v}_*\right)\right]  \,\mathrm{d}^2\mathbf{\Omega} \mathrm{d}^3\mathbf{v}_*,
\end{equation}
where $\sigma_{ss_*}$ is the differential scattering cross section. For the weak form of Eq. \eqref{elasticcollop} and other operators used in {\lb}, see Sec. \ref{vdiscsec}. Velocity space is discretized in a Burnett basis\cite{burnett_distribution_1935} with $N_k$ radial functions in velocity magnitude $v$ and $N_l^2$ spherical harmonics for velocity angles $\theta$ and $\phi$. Thus, the total number of functions in the spectral basis is $N=N_k N_l^2$. Numerical integration when required uses Gaussian quadrature with $N_v$ collocation points in $v$ or $v^2$ depending on the order of the polynomial, corresponding to the Maxwell and Laguerre basis, respectively. For calculations where the integrand is not expected to be a polynomial in $v$ (most especially, the collision operator), Gauss-Maxwell quadrature is used. Lebedev quadrature\cite{lebedev_values_1975} is used for velocity solid angles with $N_\Omega$ abscissae. More details on the velocity discretization are found in Sec. \ref{vdiscsec}. 

The spatial domain is divided into $N_z$ finite volumes, with advection fluxes computed with first-order upwinding (see Sec. \ref{spatialdisc}. Discrete time advancement is performed with forward or backward Euler, or second/fourth-order Runge--Kutta. Section \ref{timediscret} demonstrates that the conservative property is maintained for any Runge--Kutta time discretization.

\subsection{Velocity Discretization} \label{vdiscsec}

Following Refs. \cite{gamba_galerkinpetrov_2018} and \cite{kesler_fully_2019}, the Boltzmann equation is discretized with a conservative Galerkin--Petrov scheme in velocity space. After multiplying Eq. \eqref{BE} by a \emph{test function} $\psi_\alpha \left( \mathbf{v} \right)$, the distribution is expanded in an orthogonal basis $\phi_\beta \left( \mathbf{v} \right)$, we obtain the weak spectral form of the Boltzmann equation \eqref{BE}:
\begin{equation} \label{discreteBE}
   M_{\alpha \beta} \pd{f_\beta^{(s)}}{t} + T_{\alpha, \beta} \pd{f_\beta^{(s)}}{z} + F_{\alpha \beta} f_\beta^{(s)} = M_{\alpha \beta} S_\beta - L_{\alpha \beta} f_\beta^{(s)} + \sum\limits_{s_*} C_{\alpha \beta \gamma}^{(s, s_*)} f_\beta^{(s)} f_\gamma^{(s_*)},
\end{equation}
where the Einstein summation convention is implied. Greek indices are compound in the sense that $\alpha = \left(k_\alpha, l_\alpha, m_\alpha\right)$. The Burnett basis used here is defined as:
\begin{equation} \label{phidef}
   \phi_\alpha\left(\mathbf{v} \right) = \phi_{k_\alpha l_\alpha m_\alpha}\left( v, \theta, \phi \right)  = A_{k_\alpha l_\alpha} v^{l_\alpha} e^{-v^2/2 \vref^2} L_{k_\alpha}^{l_\alpha+1/2}\left(\frac{v^2}{\vref^2} \right) Y_{l_\alpha m_\alpha} \left( \theta, \phi \right).
\end{equation}
where $L_k^l$ are the associated Laguerre polynomial, $Y_{lm}$ are the real-valued spherical harmonics, and $A_{kl}$ are normalization factors to ensure orthonormality: $\int \phi_\alpha\left(\mathbf{v} \right) \phi_\beta\left(\mathbf{v} \right) \dv / \vref^3 = \delta_{\alpha \beta}$. For a given basis, a velocity scale $\vref$ is chosen, which limits the range of distributions that can be well-approximated by a limited set of basis functions.  A set of test functions that ensure conservation of the discrete collision operator are:
\begin{equation} \label{psidef}
   \psi_\alpha\left(\mathbf{v} \right) = e^{v^2/2 \vref^2} \phi_\alpha\left(\mathbf{v} \right).
\end{equation}
The following matrices in Eq. \eqref{discreteBE} are defined:
\begin{itemize}
   \item \emph{Mass matrix}
      \begin{equation} \label{massmatrix}
         M_{\alpha\beta} = \int \psi_\alpha \phi_\beta \dv
      \end{equation}
   \item \emph{Transport matrix}
      \begin{equation} \label{transportmatrix}
         T_{\alpha\beta} = \int v_z \psi_\alpha \phi_\beta \dv
      \end{equation}
   \item \emph{Force matrix}
      \begin{equation} \label{forcematrix}
         F_{\alpha\beta} = \frac{F}{m} \int \psi_\alpha \pd{\phi_\beta}{v_z}  \dv
      \end{equation}
   \item \emph{Source vector:}
      \begin{equation} \label{sourcematrix}
         S_{\alpha} = \int S \phi_\alpha \dv
      \end{equation}
   \item \emph{Loss matrix:}
      \begin{equation} \label{sourcematrix}
         L_{\alpha\beta} = \int L(\mathbf{v}) \psi_\alpha \phi_\beta \dv
      \end{equation}
\end{itemize}

The discrete collision operator ( $C_{\alpha\beta\gamma}$ in Eq. \eqref{discreteBE}) requires extended discussion for the various possible cases. For more details, see Ref. \cite{hu_fast_2019}.

First, consider the weak form form of the general discrete collision operator.
\begin{equation} \label{weakNLcoll}
   C_{\alpha\beta\gamma}^{(s,s_*)} = \int \int \int |\mathbf{v}-\mathbf{v}_*| \sigma_{ss_*}\left(|\mathbf{v} - \mathbf{v}_*|, \chi \right) B_\alpha\left(\mathbf{v},\mathbf{v}_*,\boldsymbol{\Omega}\right) \phi_\beta\left(\mathbf{v}\right) \phi_\gamma\left(\mathbf{v}_* \right) \dom \dv \dv_*,
\end{equation}
where $B_\alpha$ is a linear combination of $\psi_\alpha$ with various velocity arguments. Various special cases are listed below:
\begin{itemize}
   \item \emph{Nonlinear elastic scattering:} ($s=s_*$ only):
      \begin{equation} \label{nlB}
         B_{\alpha} =  \frac{1}{2}  \left[ \psi_\alpha\left( \mathbf{v}' \right) + \psi_\alpha\left(\mathbf{v}_*'\right) - \psi_\alpha\left( \mathbf{v} \right) - \psi_\alpha\left(\mathbf{v}_* \right)  \right]
      \end{equation}
   \item \emph{Linear elastic scattering:}
      \begin{equation}
         B_{\alpha} =   \left[ \psi_\alpha\left( \mathbf{v}' \right) - \psi_\alpha\left( \mathbf{v} \right) \right]
      \end{equation}
   \item \emph{Charge exchange:}
      \begin{equation}
         B_{\alpha} =   \left[ \psi_\alpha\left( \mathbf{v}_* \right) - \psi_\alpha\left( \mathbf{v} \right) \right]
      \end{equation}
   \item \emph{Nonlinear inelastic scattering} ($s=s_*$ only):
      \begin{equation}
         B_{\alpha} =  \frac{1}{2}  \left\{  \frac{1}{e} \left[ \psi_\alpha\left( \mathbf{v}' \right) + \psi_\alpha\left(\mathbf{v}_*'\right) \right] - \psi_\alpha\left( \mathbf{v} \right) - \psi_\alpha\left(\mathbf{v}_* \right)  \right\}
      \end{equation}
      Given a reaction with a discrete energy loss $W_\mathrm{loss}$, the coefficient of restitution is $e\left(\mathbf{v},\mathbf{v_*}\right) = \sqrt{\left(m_s v^2 + m_{s_*} v_*^2 \right)/\left(m_s v^2 + m_{s_*} v_*^2 + 2 W_\mathrm{loss} \right) }$.
   \item \emph{Linear inelastic scattering:}
      \begin{equation}
         B_{\alpha} =   \left[  \frac{1}{e}  \psi_\alpha\left( \mathbf{v}' \right) - \psi_\alpha\left( \mathbf{v} \right)  \right]
      \end{equation}

\end{itemize}
Linear collision operators against a known field particle distribution $f^{(s_*)}\left(\mathbf{w} \right)$ are obtained by likewise expanding $f^{(s_*)}$ in the same basis and applying Eq. \eqref{weakNLcoll}. 

The inherit conservation in the algorithm is encoded in Eq. \eqref{nlB}. Note that our test functions \eqref{psidef} include the functions $(1,\mathbf{v},v^2)$, whose sums are invariant under Eqs. \eqref{vprime} (they are ``collisional invariants''). Because $B_\alpha$ vanishes in Eq. \eqref{nlB} for these few $\psi_\alpha$, the collision operator exactly conserves these quantities. In practice, this is maintained to machine precision.

\subsection{Spatial Discretization} \label{spatialdisc}

Discretization in configuration space is first-order finite volume to obtain volume averages of the spectral coefficients in a basis which diagonalizes the transport matrix \eqref{transportmatrix}. Let $\mathbb{R}$ be a matrix of eigenvectors such that $\mathbb{R}\mathbb{W}\mathbb{R}^{-1} = \mathbb{M}\mathbb{T}$ yields a diagonal matrix of eigenvalues $\mathbb{W}$. The discrete Boltzmann equation for $\mathbf{g} = \mathbb{R}^{-1}\mathbf{f}$ then becomes
\begin{align} 
   \pd{g_\beta^{(s)}}{t} &+ W_\beta \pd{g_\beta^{(s)}}{z} + \left(\mathbb{R} \mathbb{M}^{-1}\right)_{\alpha\beta} F_{\beta \gamma} R^{-1}_{\gamma\delta} g_\delta^{(s)} \label{diagBE} \\
   &= R_{\alpha \beta} S_\beta - \left(\mathbb{R}\mathbb{M}^{-1}\right)_{\alpha \beta} L_{\beta \gamma} f_\gamma^{(s)} + \sum\limits_{s_*} \left(\mathbb{R} \mathbb{M}^{-1}\right)_{\alpha\beta} C_{\beta \gamma \delta}^{(s, s_*)} \left(\mathbb{R} \mathbf{g}^{(s)}\right)_\gamma \left(\mathbb{R} \mathbf{g}^{(s_*)}\right)_\delta . \nonumber
\end{align}
This transformation allows for standard upwinding techniques by defining an unambiguous advection coefficient $W_\beta$ for each spectral component of $\mathbf{g}^{(s)}$.
For further details, see Ref. \cite{kesler_fully_2019}. In a domain of length $L_z$ with $N_z$ cells that is uniform in the $x$ and $y$ direction, let the volume-averaged coefficients in the $J^\mathrm{th}$ cell 
be given by:
\begin{equation}
   g_{\alpha,J} = \frac{1}{z_{J+1/2} - z_{J-1/2}} \int\limits^{z_{J+1/2}}_{z_{J-1/2}} g_\alpha \,\mathrm{d}z.
\end{equation}
where the species superscript has been omitted for clarity, and $z_{J \pm 1/2}$ denote the boundary coordinates of the $J^\mathrm{th}$ cell. Integrate Eq. \eqref{diagBE} over this cell and apply Gauss' law:
\begin{equation}
   \pd{g_{\alpha,J}}{t} + \frac{1}{z_{J+1/2} - z_{J-1/2}} \left( \Gamma_{\alpha,J+1/2} - \Gamma_{\alpha,J-1/2} \right) = Z_\alpha,
\end{equation}
where $\Gamma_{\alpha,J\pm1/2}$ are the fluxes on either side of cell $J$ and $Z_\alpha$ represents every other term from Eq. \eqref{diagBE}.
The fluxes are calculated from $\mathbf{g}_\alpha$ in a way that recreates the upwind finite difference method:
\begin{align} \label{upwinding}
   \Gamma_{\alpha,J+1/2} &= \frac{W_\alpha}{2}\left( g_{\alpha,J} + g_{\alpha,J+1} \right) + \frac{|W_\alpha|}{2} \left( g_{\alpha,J} - g_{\alpha,J+1} \right) \nonumber \\
   \Gamma_{\alpha,J-1/2} &= \frac{W_\alpha}{2}\left( g_{\alpha,J-1} + g_{\alpha,J} \right) + \frac{|W_\alpha|}{2} \left( g_{\alpha,J-1} - g_{\alpha,J} \right).
\end{align}
At the boundaries of the domain ($J=1/2,N_z+1/2$ corresponding to $z=0,L_z$ respectively), the fluxes, as projected onto the diagonalized basis, are specified as boundary conditions. Alternatively, the array of spectral coefficients $\mathbf{f}$ (and thereby $\mathbf{g}$) can be specified as Dirichlet boundary conditions. In addition to a diffusively reflecting boundary (see Ref. \cite{kesler_fully_2019}), a specularly reflecting boundary is implemented from the property of spherical harmonics upon reflection across the $v_z = 0$ plane:
\begin{equation}
   g^{+}_{\alpha,J} =  \left(-1 \right)^{l_\alpha+m_\alpha} g^{-}_{\alpha,J},
\end{equation}
where $g^{-}_{\alpha,J}$ are the cell-averaged spectral coefficients in the cells adjacent to the boundaries as estimated from the previous timestep, and $g^+_{\alpha,J}$ are those which are used in Eq. \eqref{upwinding} to calculate the flux at the boundaries.

\subsection{Time Discretization} \label{timediscret}

Temporal discretization can be performed with forward/backward Euler or second/fourth order Runge--Kutta. For implicit advance of the nonlinenar collision operator, Newton's method is used on the problem's Jacobian, reducing to the predictor-corrector method when a single Newton iteration is used. 

In the following, we analyze the conservation properties of these schemes.
To that end, we consider the semi-discrete Boltzmann equation without transport and force terms
as these usually only conserve mass:
\begin{equation}\label{eq:boltzmann-homogeneous}
M_{\alpha \beta} \pd{f_\beta}{t}
= C_{\alpha \beta \gamma} f_\beta f_\gamma.
\end{equation}
To simplify notation, we focus on the case of a single particles species.
The results generalize immediately to multiple species.
For a coefficient vector $\mathbf h$ we define its moment bracket $\langle \mathbf h \rangle_{\mathfrak m}$ by
\begin{equation}
\langle \mathbf h \rangle_{\mathfrak m} = \int_{\mathbb R^3} f_\beta \varphi_\beta(\mathbf v)
\begin{pmatrix}
\psi_{\alpha_1}(\mathbf v) \\ \vdots \\ \psi_{\alpha_5}(\mathbf v)
\end{pmatrix}
\, \text d \mathbf v.
\end{equation}
for the five collision invariants $\psi_{\alpha_1},\dots,\psi_{\alpha_5}$.
The moment bracket can be cast in algebraic form,
\begin{equation}
\langle \mathbf h \rangle_{\mathfrak m}
= P_{i \alpha} M_{\alpha \beta} h_\beta
\end{equation}
with $P_{i \alpha} = 1$ if $\alpha = \alpha_i$
and $0$ otherwise.
For the solution $\mathbf f$ of Eq.~\eqref{eq:boltzmann-homogeneous}
it holds
\[
\frac{\text d}{\text d t} \langle \mathbf f \rangle_{\mathfrak m}
= P_{i \alpha} M_{\alpha \beta} \pd{f_\beta}{t}
= P_{i \alpha} C_{\alpha \beta \gamma} f_\beta f_\gamma
= 0
\]
since the collision matrices for $\alpha_1,\dots,\alpha_5$ vanish
owing to the choice of the test functions.
This conservation property is preserved if the system of differential equations
is discretized by \emph{any} Runge--Kutta method.
To prove this, we consider a single time step of size $\tau > 0$
with an $m$-stage Runge--Kutta method,
\begin{equation}
\mathbf f^{N + 1} = \mathbf f^{N} + \tau \sum_{i = 1}^m b_i \mathbf k_i,
\end{equation}
where the vectors $\mathbf k_1, \dots, \mathbf k_m$ are solutions to
\begin{equation}\label{eq:runge-kutta-k}
\mathbf k_i = \mathcal C \Big(
\mathbf f^{N} + \tau \sum_{\ell = 1}^m a_{j \ell} \mathbf k_\ell
\Big), \quad i = 1, \dots, m,
\end{equation}
for coefficients $b_1, \dots, b_m$, a matrix $(a_{i j})_{i, j = 1}^m$
and
\begin{equation}
\mathcal C(\mathbf h)_{\alpha}
= M^{-1}_{\alpha \beta} C_{\beta \gamma \delta} h_{\gamma} h_\delta.
\end{equation}
For the method to be conservative we need
\begin{equation}
\langle \mathbf f^{N + 1} \rangle_{\mathfrak m}
= \langle \mathbf f^{N} \rangle_{\mathfrak m},
\end{equation}
for which a sufficient condition is
\begin{equation}
\langle \mathbf k_j \rangle_{\mathfrak m} = 0, \quad j=1,\dots,m.
\end{equation}
This, however, is a direct consequence of the algebraic form of the moment bracket:
\begin{equation}
\langle \mathbf k_j \rangle_{\mathfrak m}
= P_{i \alpha} M_{\alpha \beta}
\mathcal C\Big(
\mathbf f^{N} + \tau \sum_{\ell = 1}^m a_{j \ell} \mathbf k_\ell
\Big)_\beta
=
P_{i \alpha} M_{\alpha \beta}
M^{-1}_{\beta \gamma} C_{\gamma \delta \varepsilon}
h_\delta h_\varepsilon,
\end{equation}
where we write $\mathbf h$ for the argument of $\mathcal C$.
This expression simplifies to
\[
\langle \mathbf k_j \rangle_{\mathfrak m}
= P_{i \alpha} C_{\alpha \delta \varepsilon} h_\delta h_\varepsilon
= 0
\]
by the same argument as in the continuous case.

Above result shows that every explicit Runge--Kutta method
conserves mass, momentum and energy.
For an implicit treatment of the collision operator it also conveys
that conservation is always preserved on the discrete level,
regardless of the method
used for the (approximate) solution
of the nonlinear system~\eqref{eq:runge-kutta-k}.
For a fixed-point or Newton's method an even stronger statement holds true:
Any intermediate iteration for the solution of~\eqref{eq:runge-kutta-k}
conserves the collision invariants.
The proof of this statement only requires minor modifications of above arguments.


\section{Applications and Benchmarks} \label{benchmarksec}

The Galerkin--Petrov method for the Boltzmann collision operator has been extensively benchmarked against the constructed solution of Maxwell pseudomolecules in Refs. \cite{gamba_galerkinpetrov_2018} and \cite{kesler_fully_2019}. {\lb}  reproduces the success of these tests in addition to an inelastic generalization \cite{hu_fast_2019}, wherein a constant restitution coefficient exponentially decreases the temperature of the gas. Despite an order of magnitude drop in the temperature, the reference Maxwellian at a resolution of $N=27$ is able to faithfully capture to the transient temperature drop with a relative error of $3.6\times 10^{-4}$.

In this section, we build upon these tests ensuring that it remains robust and accurate for more general cross sections, spatial advection, acceleration, and source/sink terms. This is done with a combination of analytic results and comparing to other established Boltzmann solvers in their appropriate limits. Table \ref{benchmarktable} summarizes the physics tested in each benchmark.

\begin{table}
   \caption{\label{benchmarktable} A summary of the analytic and cross-code benchmarks for {\lb}. Vertically listed are the various terms in the Boltzmann equation. A check mark under a column indicates that term was an important component the corresponding benchmark.}
\begin{center}
   \begin{tabular}{l | c c c c c|}
      Term in Boltzmann equation & Maxwell molecules & Chapman-Enskog & \textsc{bolsig+} & \textsc{degas2} \\
      \hline
      Transience & \checkmark & & &  \\
      Spatial advection & & \checkmark  & & \checkmark \\
      Force-field acceleration & & & \checkmark & \\
      Nonlinear collision operator & \checkmark & \checkmark & & \\
      Linear collision operator & & & \checkmark & \checkmark \\
      Inelastic scattering & \checkmark & & & \\
      Sources, boundary conditions & & & & \checkmark \\
   \end{tabular}
\end{center}
\end{table}

\subsection{Chapman-Enskog expansion} \label{cesec}

One of the most successful applications of the Boltzmann equation is the Chapman-Enskog expansion, which leads to the Navier-Stokes equations for neutral fluids. It is an expansion for small Knudsen number $\mathrm{Kn} = \lambda / l \ll 1$, where $l$ is a representative length scale, $\lambda = \left( \sigma_0 n \right)^{-1}$ is the  \emph{mean free path}, $\sigma_0$ is a representative collision cross section, and $n$ the target number density. In this limit, the gas is highly collisional relative to other scales of interest, thereby remaining close to a Maxwell-Boltzmann distribution in velocity.

A key result of this calculation\cite{chapman_s._c._mathematical_1990} is that the heirarchy of moment equations can be closed by the following relations for the momentum and heat flux, respectively:
\begin{equation} \label{stresstensor}
   \Pi_{ij} = \frac{m}{3} \int \left(v_i - U_i \right) \left(v_j - U_j \right) f\left(\mathbf{v} \right) \,\mathrm{d}^3\mathbf{v} = - n m \nu_{CE} \left( \pd{U_i}{x_j} + \pd{U_j}{x_i} - \frac{2}{3} \delta_{ij} \nabla \cdot \mathbf{U} \right) 
\end{equation}
\begin{equation} \label{hflux}
   q_i = \frac{m}{2} \int \left(v_i - U_i \right) | \mathbf{v} - \mathbf{U} |^2 f\left(\mathbf{v} \right) \,\mathrm{d}^3\mathbf{v} = - n \chi_{CE} \pd{T}{x_i}.
\end{equation}
The density, flux, and scalar pressure are defined as moments of the distribution function:
\begin{equation} \label{basicmoments}
   n = \int f(\mathbf{v}) \dv, \,\,\,\,\,\,\, \boldsymbol{\Gamma} = \int \mathbf{v} f(\mathbf{v}) \dv, \,\,\,\,\,\,\, p =  \frac{m}{3} \int | \mathbf{v} - \mathbf{U} |^2 f(\mathbf{v}) \dv
\end{equation}
while flow velocity and temperature are $\mathbf{U} = \boldsymbol{\Gamma}/n$ and $T=p/n$, respectively.
For a gas of elastic hard spheres with a constant cross section, $\sigma = \sigma_0$, Chapman-Enskog theory provides asymptotic estimates for the viscosity\footnote{In Eq. \eqref{stresstensor} the \emph{second viscosity}, associated with expansion, has been omitted because it is not relevant in the 1D test cases shown here.}:
\begin{equation} \label{ce_visc}
   \nu_{CE} = 1.016 \frac{5 \sqrt{\pi}}{16 \sigma_0 n} \sqrt{\frac{T}{m}},
\end{equation}
and thermal diffusivity
\begin{equation} \label{ce_diff}
   \chi_{CE} = 1.02513 \frac{75 \sqrt{\pi}}{64 \sigma_0 n} \sqrt{\frac{T}{m}}.
\end{equation}
Since {\lb} solves the full Boltzmann equation, it ought to reproduce this theory in the highly collisional limit, even at low velocity resolution. The fluid closure \eqref{stresstensor}-\eqref{hflux} is expressed in terms of gradients, so a spatial domain is needed to compare the results of {\lb} to theory in the collisional limit.

A 1D spatial domain is simulated with diffusively-reflecting boundary conditions at different wall temperatures and velocities (corresponding to the Fourier and Couette problems, respectively). After {\lb} evolves the Boltzmann equation to steady-state, the temperature and velocity gradients are measured, along with the respective predicted fluxes of heat and momentum. The number of spectral coefficients kept for this benchmark corresponds to $N_k = 3$ and $N_l = 3$, while the velocity space integrations use $N_v = 8$ and $N_\Omega = 38$. The timestep is 1 $\mu\mathrm{s}$ and the steady-state was measured at 50 ms. The domain is 40m in length, split into 16 discrete finite volumes.

\begin{figure}
   \begin{center}
   \includegraphics[width=0.4\textwidth]{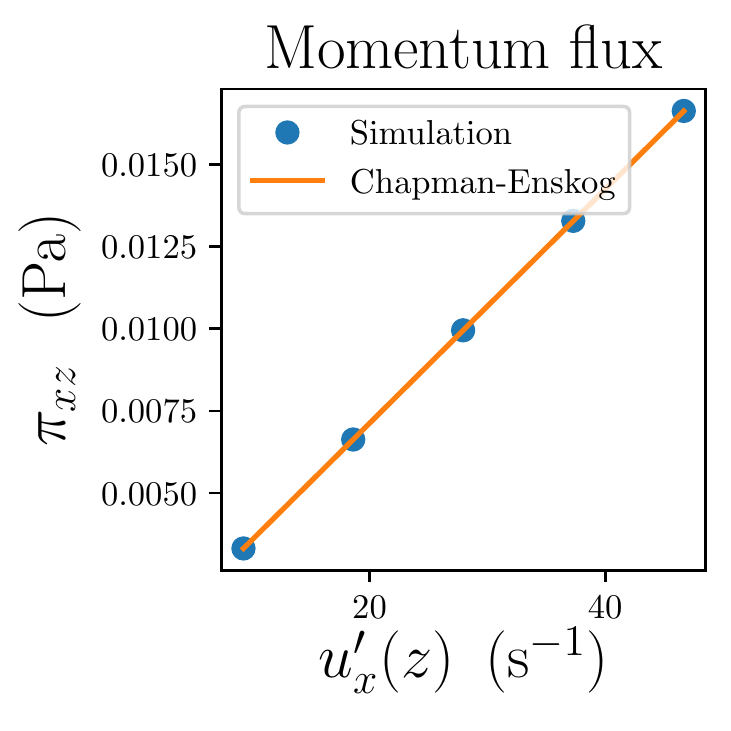}
   \includegraphics[width=0.4\textwidth]{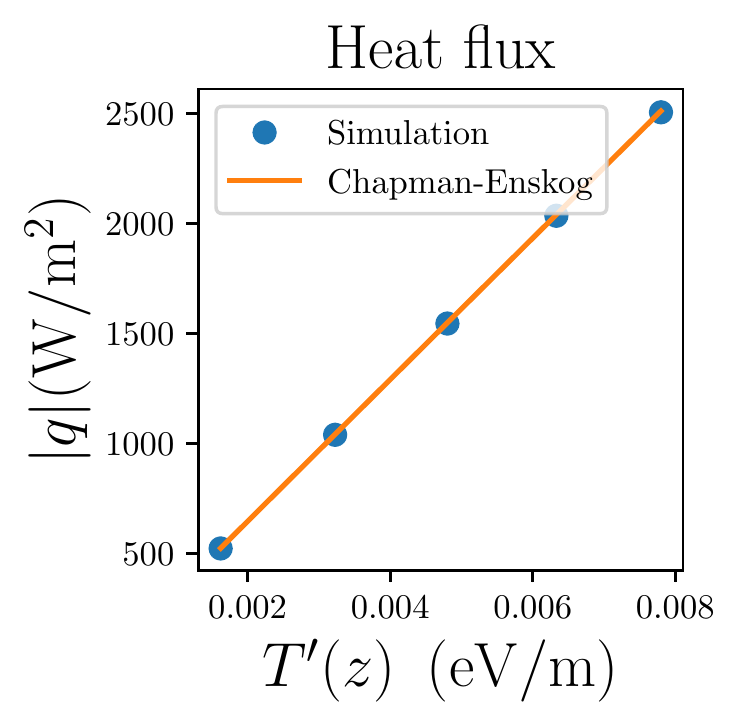}
   \end{center}
   \caption{\label{cecomp} Comparing the momentum (left) and heat (right) fluxes as predicted by {\lb} and the Chapman-Enskog prediction at $\mathrm{Kn} = 0.01$. In the latter, corrections up to $\mathcal{O} \left(\mathrm{Kn}^3 \right)$ have been retained as reflected in Eqs. \eqref{ce_visc} \eqref{ce_diff}. }
\end{figure}

Figure \ref{cecomp} shows that {\lb} reproduces well the analytic Chapman-Enskog calculations for a nominal $\mathrm{Kn}=0.01$. With next-order finite-Kn corrections to the transport coefficients $\chi$ and $\nu$, {\lb} agrees with the theory up to 3-5 significant figures. As the gradients increase such that the Knudsen number approaches unity, the results begin to diverge, reflecting the expected inaccuracy in the Chapman-Enskog expansion.
At large Knudsen number, the interior distribution is more strongly coupled to the boundary conditions, which were taken here to be Maxwellian. As such, the viscosity and diffusivity become increasingly meaningless as the gas becomes less collisional.

\subsection{Neutrals in magnetic confinement fusion}

The non-confined region of magnetic confinement fusion experiments is known as the ``scrape off layer'' (SOL). It is here that the plasma leaked from the confined region makes its way to the solid wall. The dynamics of the SOL are critical in determining the viability of a fusion power plant since tremendous energy can be deposited over a small area, causing impurities to be ejected from the wall among other deleterious effects. As a means of mitigation, experiments can be designed to cool the plasma before it reaches the wall, even to the point of recombination into neutral atoms.
In the extreme case, a significant population of neutral gas absorbs plasma energy in a way that is emitted by recombined atoms via line radiation. This process is known as \emph{detachment}, and the dynamics of neutrals in the fluid, transition, and free-molecular flow regimes are critical to accurate prediction of the particle, momentum, and power balance in the SOL~\cite{krasheninnikov_divertor_2016,stangeby_basic_2018}.


Nonlinear neutral-neutral scattering has been shown to be an important local effect in detached divertor conditions owing to the high neutral density relative to the plasma\cite{kukushkin_effect_2005,kotov_numerical_2007}.
\textsc{degas2}\cite{stotler_neutral_1994} and \textsc{eirene}\cite{reiter_randschicht-konfiguration_1984} are state-of-the art Monte Carlo Boltzmann solvers to predict the behavior of neutral atoms and molecules and their effect on the plasma. The nonlinear elastic scattering process among neutral species is modelled in these codes as a simple Bhatnagar-Gross-Krook (BGK) operator\cite{bhatnagar_model_nodate}, whose nonlinear character is respected by iterating toward convergence the Maxwellian to which the BGK operator relaxes. The collision frequency and its velocity dependence is determined by matching to the viscosity in the fluid limit (see section \ref{cesec}). In this way, although these tools use a simplified nonlinear collision operator, they still reproduce the fluid result of the Couette problem\cite{kotov_numerical_2007}. They do not, however, capture the correct thermal diffusivity in the fluid limit and cannot be expected to evaluate collisions properly in the transitional Knudsen regime.

In this section, {\lb} is benchmarked against \textsc{degas2} in a regime where both codes are expected to produce equivalent results: a 1D domain of neutral atoms scattering off of a fixed plasma background that varies sharply in space. The test case represents a 1D model of the conditions in a tokamak scrape-off layer. The domain has spatial extent $L_z = 2 \mathrm{m}$. The source is injected at a rate of $10^{24}$ hydrogen atoms per $\mathrm{m}^2$ per s inward at the domain boundary. The velocity dependence of the flux is proportional to $\exp\left(-m v^2/2T_w\right)$. The ``wall'' temperature is $T_w = 3 \mathrm{eV}$, approximately representing atoms produced from dissociating hydrogen molecules. The plasma has a uniform density of $n_e = n_i = 5\times10^{19} \mathrm{m}^{-3}$ and a temperature dependence given by:
\begin{equation}
   T_{e/i}(z) = T_w + (T_{e/i,up} - T_w) \tanh\left( \frac{L_z - |L_z - 2 z| }{\Delta} \right)
\end{equation}
where $\Delta = 0.1 \mathrm{m}$ and the upstream electron and ion temperatures are $T_{e,up} = 10 \mathrm{eV}$ and $T_{i,up} = 20 \mathrm{eV}$, respectively. Both simulations include the following hydrogen reactions: charge exchange\cite{krstic_elastic_1998}, electron-impact ionization, and plasma recombination (the latter two as modified by a collisional radiative model\cite{janev_elementary_1988,bray_electron_2012}). {\lb} uses the same atomic data tables for these reactions as \textsc{degas2}. Particles leaving the domain reflect specularly from the boundary with a reflection coefficient of 0.9. This is similar to a recent benchmark comparison of \textsc{degas2} with a recent expansion of the \textsc{gkeyll} framework\cite{bernard_kinetic_2022}.

\begin{figure}
   \includegraphics[width=1.0\textwidth]{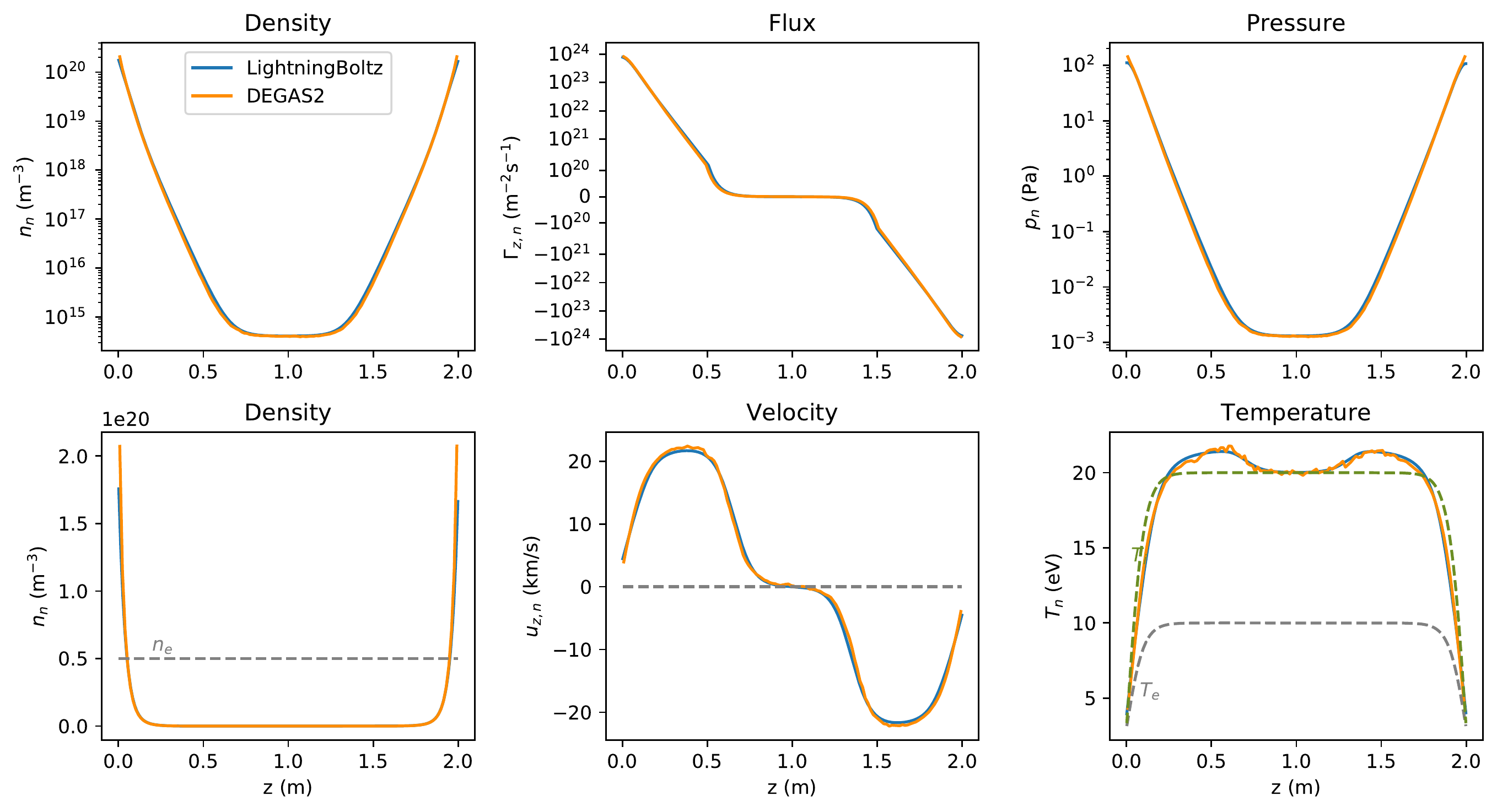}
   \caption{\label{d2comp}Comparing the results for {\lb} with the \textsc{degas2} Monte Carlo solver for neutrals in a 1D scrape-off-layer-like domain. Superimposed on the lower plots are the density and temperatures of the plasma.}
\end{figure}

{\lb} is run transiently to a steady-state, while \textsc{degas2} calculates the strict steady-state using Monte Carlo tally estimation. The \textsc{degas2} results shown here use 2M trajectory samples, using track-length estimators for suppressed ionization\cite{spanier_two_1966}. {\lb} uses a spectral velocity discretization with $N_l = 3$ and $N_k=3$ and a reference temperature of 5\,eV throughout the domain. The comparison is shown is Fig. \ref{d2comp}.

Both codes strive to calculate the distribution function and/or moments thereof as spatial averages within the finite volumes as discretized. With track-length estimators, \textsc{degas2} calculates the spatial advection and ionization loss \emph{analytically} per trajectory sample. So the finite volume resolution does not affect the resolution of advection in \textsc{degas2}, while it does in {\lb}. Therefore, for the results presented here,  $N_z = 128$ in \textsc{degas2} and $N_z = 512$ in {\lb}.
 Disagreement at spatial resolution lower than this in {\lb} in is most apparent in the low density region: around $z$=0.5 m. Even when only 128 finite volumes are used in {\lb}, however, agreement within about 3\% is observed near the wall ($z < 0.2 \mathrm{m}$). In the \textsc{degas2} simulations, significant noise is observed near the middle of the domain. This is expected since velocity and temperature are \emph{ratios} of moments which are directly calculated, each with some finite-sampling noise. Since the neutral density is quite low in the interior region, this amplifies the noise present, but this also means that the physical consequence of this noise is not particularly troubling. 

\begin{figure}
    \centering
    \includegraphics[width=1.0\textwidth]{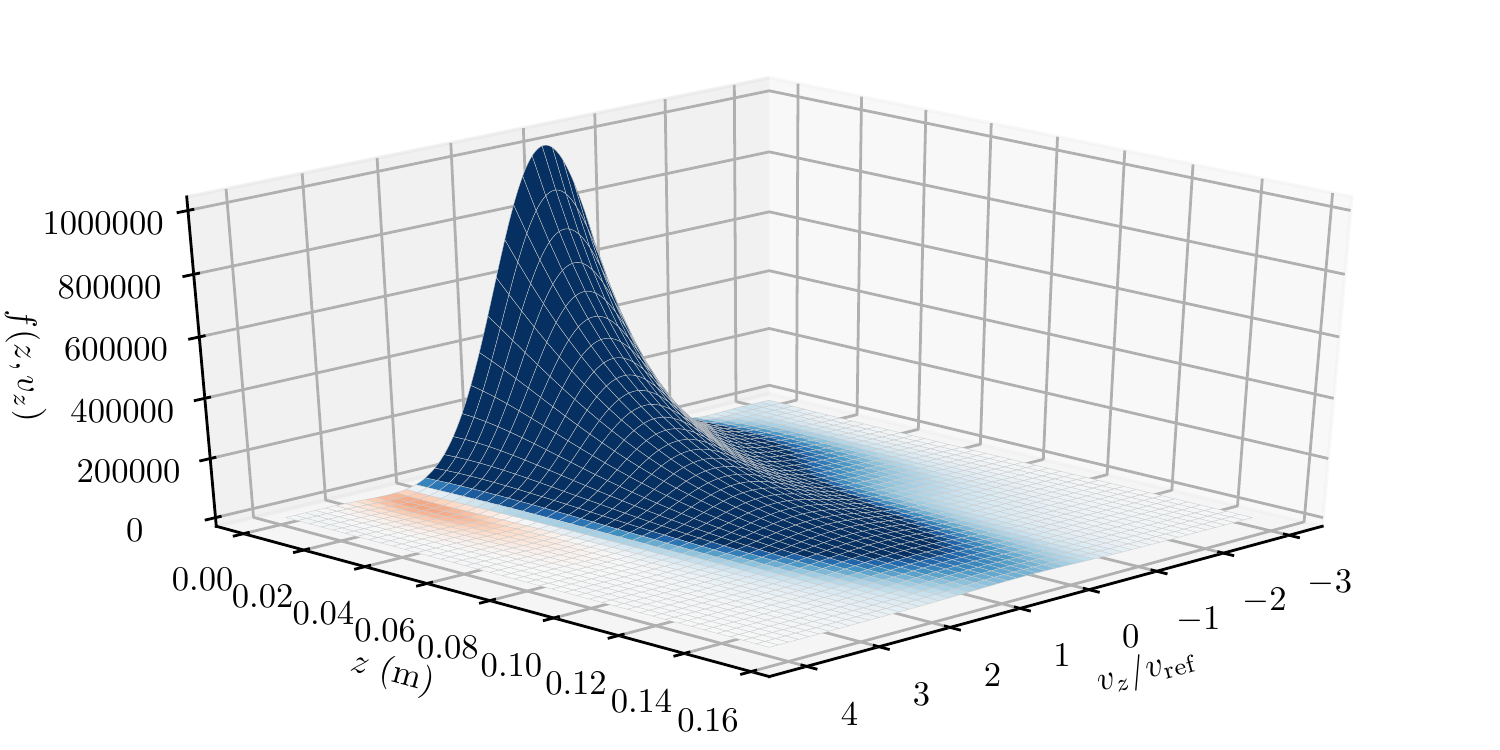}
    \caption{The distribution function calculated by {\lb} for the \textsc{degas2} benchmark (Fig. \ref{d2comp}) as projected from the orthogonal basis onto the $(z-v_z)$ plane at $v_x=v_y=0$. Colors are such that the red is negative, white is zero, and blue is positive.}
    \label{fvz}
\end{figure}

Figure \ref{d2comp} shows that very good agreement is found despite the two solvers taking very different approaches to solving the Boltzmann equation.
The fixed ion distribution is expanded in the same basis as the neutrals although the plasma temperature varies by an order of magnitude. The neutrals closely follow the ion temperature in this case due to the charge exchange process, so at this relatively low spectral resolution, one cannot expect the distribution function to be well-behaved when projected back to a function of $\mathbf{v}$. See Fig. \ref{fvz} for such a projection.
In particular, $f^{(n)}(\mathbf{v})$ is observed to be oscillatory and negative in some regions of phase-space. This is partly a consequence of solving the \emph{weak form} of the Boltzmann equation: even at this low resolution in {\lb}, the physically relevant moments examined are well captured. 
The unphysical behavior can be mitigated by using a higher spectral resolution, by allowing $T_\mathrm{ref}$ to be a function of $z$ \cite{kesler_fully_2019}, or appealing to the concept of \emph{weak positivity}, wherein there may exist \emph{a} positive-definite distribution that shares the same set of moments solved by the weak form\cite{hakim2020continuum,mandell_magnetic_2020}. Such a property has not yet been demonstrated for the Boltzmann Galerkin--Petrov algorithm.



\subsection{Weakly ionized plasma} \label{bolsigbenchmarksec}

Spectral expansions have long been used to find approximate solutions to the Boltzmann equation for electrons scattering in weakly ionized plasmas. The archetypal method is the so-called two-term approximation, wherein the electron distribution function is assumed to be only weakly anisotropic. The expansion in spherical harmonics is kept only up to only up to first order with one azimuthal mode, and the energy distribution of the anisotropic perturbation (which carries the current) is solved by conventional means. \textsc{bolsig+}\cite{hagelaar_solving_2005} is a flagship code that performs this calculation efficiently and can be freely run online\cite{hagelaar_bolsig_nodate}.

In this section, we compare steady-state swarm parameters calculated by {\lb} to that from \textsc{bolsig+}. We choose electrons elastically scattering off of ground state Argon atoms\cite{alves_ist-lisbon_2014,pitchford_lxcat_2017} while accelerated by constant electric fields $E$ ranging from 0.01-0.1 V/m. 
In {\lb}, we likewise only expand the distribution function to $N_l=2$, and the energy spectra are resolved up to $N_k = 7$. A reference temperature of $0.5 \mathrm{eV}$ was used for the basis in all cases. 

\begin{figure}
   \begin{center}
   \includegraphics[width=1.0\textwidth]{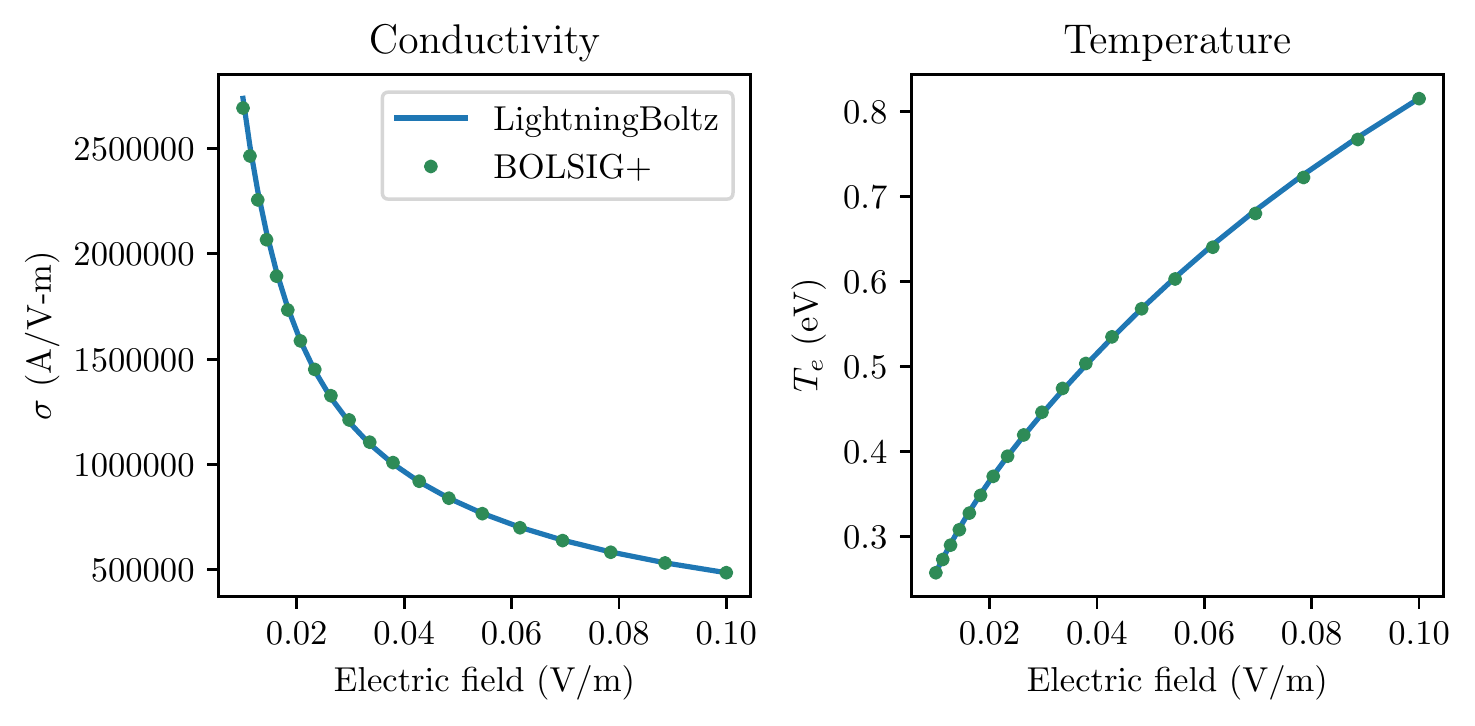}
   \end{center}
   \caption{\label{bolsigcomp} Comparing the results of {\lb} with \textsc{bolsig+} under the two-term approximation for electrons scattering elastically off of argon.}
\end{figure}

The conductivity is defined by $J = - e n_e U_z = \sigma E$ (with elementary charge $e$), motivated by the expectation that in steady-state, the current density $J$ will be proportional to the electric field. Likewise, in a steady state the electrons will settle to a distribution with a particular temperature as the energy gained by Joule heating is balanced with the energy lost by scattering from the Argon atoms. The comparison of these quantities reported by the respective codes\footnote{\textsc{bolsig+} actually reports the electron mobility and average kinetic energy, which are directly related to conductivity and temperature, respectively.}  is shown in Fig. \ref{bolsigcomp}. Again, although the temperature departs significantly from the reference energy, these moments remain well estimated.

\section{Discussion and Conclusion} \label{discsec}

The generality and difficulty of solving the nonlinear Boltzmann equation demands a general-purpose solver to test assumptions, derive reduced models, or perform direct simulation of physical systems. In this work, the suitability of the Galerkin--Petrov method as implemented in {\lb} has been demonstrated for a broad range of physical problems (Table \ref{benchmarktable}). Most of these benchmarks compared \emph{moments} of kinetic solutions, which is appropriate when solving the weak form of the Boltzmann equation.

For cases that demand accuracy of the distribution function $f(\mathbf{v})$ itself, accurate representation of moments is not sufficient. In particular, consider wave-particle interactions such as Landau damping, which are sensitive to velocity space gradients of the distribution function. In this work, only modest velocity space resolution was required, but this is by no means universal. At sufficiently high spectral resolution, the Galerkin-Petrov method is capable of fully resolving highly non-Maxwellian distributions at sufficiently high spectral resolution (see Ref. \cite{kesler_fully_2019}). When too few spectral coefficients are used, Gibbs' phenomenon results in unphysical oscillations and negativity of the distribution function (see Fig. \ref{fvz}). Even when this is the case, the moments are well behaved as long as the integrals are evaluated accurately. Perhaps surprisingly, this is even the case when the cross section is a nontrivial functions of velocity, which one may expect to be sensitive to unphysical features in the distribution. 

In most cases, inaccuracy of the limited spectral representation of the distribution function can be attributed to disparate energy scales. The function about which the basis is constructed may be a poor representation of the local distribution function.
This can be mitigated by allowing the reference temperature to vary in space~\cite{kitzler_high_2015}. The case of extreme velocity-space anisotropy, such as that of relativistic runaway electrons, is more troublesome. If one velocity angle is ignorable (as is often the case in magnetized plasma), the 2D symmetries of spherical harmonics are not as relevant and an alternate basis in $\theta$ could be constructed.

Considering the $N^3$ complexity of solving the Boltzmann equation, a spectral resolution of $\sim 800$ coefficients is expected to be the limit for advancing the collision operator when configuration space is parallelized into shared-memory nodes. It is those applications which require high spectral resolution that will benefit strongly from the distributed precomputation of {\lb}. The integration method for this precomputation can be adapted to the problem at-hand. For example, in the case of electrons in weakly-ionized plasma, wherein the cross section is independent of the target velocity, a quadrature method weighted by the cross section itself can be constructed \cite{wilkie_microturbulent_2015}, and this is currently implemented in {\lb}.

In addition, several applications expect an equilibrium distribution which is not Maxwellian, and such a distribution may be a more suitable function around which to build a basis. A classic example is the  Druyvesteyn distribution \cite{druyvesteyn_influence_1930,morse_velocity_1935} commonly used in weakly ionized plasma applications and is rigorous for simplified elastic cross sections and strong electric field. The spectral method can straightforwardly be extended in {\lb} to a basis defined with this measure rather than the Maxwellian distribution. As long as the same test functions \eqref{psidef} are used, the algorithm will maintain its conservative properties even in an alternate basis.

The authors are appreciative to Irene Gamba, Daren Stotler, Felix Parra, Greg Hammett, and T\"unde F\"ul\"op for their support, expert guidance, and fruitful discussions. This work was supported by the United States Department of Energy via contract no. DE-AC02-09CH11466, and the SciDAC Center for High-Fidelity Boundary Plasma Simulation.

\bibliographystyle{unsrt}
\bibliography{zotero}

\begin{thebibliography}{10}

\bibitem{gamba_galerkinpetrov_2018}
I.~M. Gamba and S.~Rjasanow.
\newblock Galerkin{\textendash}{Petrov} approach for the {Boltzmann} equation.
\newblock {\em Journal of Computational Physics}, 366:341--365, August 2018.

\bibitem{kesler_fully_2019}
T.~Ke{\ss}ler and S.~Rjasanow.
\newblock Fully conservative spectral {Galerkin}{\textendash}{Petrov} method
  for the inhomogeneous {Boltzmann} equation.
\newblock {\em Kinetic \& Related Models}, 12(3):507--549, 2019.

\bibitem{landau_kinetic_1936}
L.~D. Landau.
\newblock Kinetic equation for the {Coulomb} effect.
\newblock {\em Phys. Z. Sowjetunion}, 10:154, 1936.

\bibitem{braginskii_transport_1965}
S.~I. Braginskii.
\newblock Transport processes in a plasma.
\newblock In M.~A. Leontovich, editor, {\em Reviews of {Plasma} {Physics}},
  volume~1, page 205. Consultants Bureau, New York, 1965.

\bibitem{chapman_s._c._mathematical_1990}
S.~C. Chapman and T.~G. Cowling.
\newblock {\em The mathematical theory of non-uniform gases: an account of the
  kinetic theory of viscosity, thermal conduction and diffusion in gases}.
\newblock Cambridge University Press, 1990.

\bibitem{ziman_electrons_2001}
J.~M. Ziman.
\newblock {\em Electrons and phonons: the theory of transport phenomena in
  solids}.
\newblock Oxford University Press, 2001.

\bibitem{drewes_boltzmann_2013}
M.~Drewes, S.~Mendizabal, and C.~Weniger.
\newblock The {Boltzmann} equation from quantum field theory.
\newblock {\em Physics Letters B}, 718(3):1119--1124, January 2013.

\bibitem{bird_molecular_1994}
G.~A Bird.
\newblock {\em Molecular gas dynamics and the direct simulation of gas flows}.
\newblock Oxford University Press, Oxford, 1994.

\bibitem{nanbu_direct_1980}
K.~Nanbu.
\newblock Direct {Simulation} {Scheme} {Derived} from the {Boltzmann} equation.
  {I}. {Monocomponent} {Gases}.
\newblock {\em Journal of the Physical Society of Japan}, 49(5):2042, 1980.

\bibitem{galitzine_accuracy_2014}
C.~Galitzine.
\newblock {\em On the {Accuracy} and {Efficiency} of the {Direct} {Simulation}
  {Monte} {Carlo} {Method}}.
\newblock PhD thesis, University of Michigan, 2014.

\bibitem{bird_forty_2001}
G.~A. Bird.
\newblock Forty years of {DSMC}, and now?
\newblock In {\em {AIP} {Conference} {Proceedings}}, volume 585, pages
  372--380, Sydney (Australia), 2001. AIP.

\bibitem{tantos_deterministic_2020}
C.~Tantos, S.~Varoutis, and C.~Day.
\newblock Deterministic and stochastic modeling of rarefied gas flows in fusion
  particle exhaust systems.
\newblock {\em Journal of Vacuum Science \& Technology B}, 38(6):064201,
  November 2020.

\bibitem{joseph_coupling_2017}
I.~Joseph, M.~E. Rensink, D.~P. Stotler, A.~M. Dimits, L.~L. LoDestro, G.~D.
  Porter, T.~D. Rognlien, B.~Sjogreen, and M.~V. Umansky.
\newblock On coupling fluid plasma and kinetic neutral physics models.
\newblock {\em Nuclear Materials and Energy}, 12:813--818, August 2017.

\bibitem{juno_discontinuous_2018}
J.~Juno, A.~Hakim, J.~TenBarge, E.~Shi, and W.~Dorland.
\newblock Discontinuous {Galerkin} algorithms for fully kinetic plasmas.
\newblock {\em Journal of Computational Physics}, 353:110--147, January 2018.

\bibitem{pitchford_extended_1981}
L.~C. Pitchford, S.~V. ONeil, and J.~R. Rumble.
\newblock Extended {Boltzmann} analysis of electron swarm experiments.
\newblock {\em Physical Review A}, 23(1):294--304, January 1981.

\bibitem{yachi_multi-term_1988}
S.~Yachi, Y.~Kitamura, K.~Kitamori, and H.~Tagashira.
\newblock A multi-term {Boltzmann} equation analysis of electron swarms in
  gases.
\newblock {\em Journal of Physics D: Applied Physics}, 21(6):914--921, June
  1988.

\bibitem{pareschi_fourier_1996}
L.~Pareschi and B.~Perthame.
\newblock A {Fourier} spectral method for homogeneous boltzmann equations.
\newblock {\em Transport Theory and Statistical Physics}, 25(3-5):369--382,
  April 1996.

\bibitem{filbet_solving_2006}
F.~Filbet, C.~Mouhot, and L.~Pareschi.
\newblock Solving the {Boltzmann} {Equation} in \textit{{N}} log
  $_{\textrm{2}}$ \textit{{N}}.
\newblock {\em SIAM Journal on Scientific Computing}, 28(3):1029--1053, January
  2006.

\bibitem{filbet_analysis_2011}
F.~Filbet and C.~Mouhot.
\newblock Analysis of spectral methods for the homogeneous {Boltzmann}
  equation.
\newblock {\em Transactions of the American Mathematical Society},
  363(04):1947--1947, April 2011.

\bibitem{burnett_distribution_1936}
D.~Burnett.
\newblock The distribution of molecular velocities and the mean motion in a
  non-uniform gas.
\newblock {\em Proceedings of the London Mathematical Society}, 2(1):382--435,
  1936.
\newblock Publisher: Wiley Online Library.

\bibitem{bezanson_julia_2017}
J.~Bezanson, A.~Edelman, S.~Karpinski, and V.~B. Shah.
\newblock Julia: {A} fresh approach to numerical computing.
\newblock {\em SIAM review}, 59(1):65--98, 2017.
\newblock Publisher: SIAM.

\bibitem{hu_fast_2019}
J.~Hu and Z.~Ma.
\newblock A fast spectral method for the inelastic {Boltzmann} collision
  operator and application to heated granular gases.
\newblock {\em Journal of Computational Physics}, 385:119--134, May 2019.

\bibitem{burnett_distribution_1935}
D.~Burnett.
\newblock The distribution of velocities in a slightly non-uniform gas.
\newblock {\em Proceedings of the London Mathematical Society}, 2(1):385--430,
  1935.
\newblock Publisher: Wiley Online Library.

\bibitem{lebedev_values_1975}
V.~I. Lebedev.
\newblock Values of the nodes and weights of ninth to seventeenth order
  gauss-markov quadrature formulae invariant under the octahedron group with
  inversion.
\newblock {\em USSR Computational Mathematics and Mathematical Physics},
  15(1):44--51, 1975.

\bibitem{krasheninnikov_divertor_2016}
S.~I. Krasheninnikov, A.~S. Kukushkin, and A.~A. Pshenov.
\newblock Divertor plasma detachment.
\newblock {\em Physics of Plasmas}, 23(5):055602, May 2016.

\bibitem{stangeby_basic_2018}
P.~C. Stangeby.
\newblock Basic physical processes and reduced models for plasma detachment.
\newblock {\em Plasma Physics and Controlled Fusion}, 60(4):044022, April 2018.

\bibitem{kukushkin_effect_2005}
A.~S. Kukushkin, H.~D. Pacher, V.~Kotov, D.~Reiter, D.~Coster, and G.~W.
  Pacher.
\newblock Effect of neutral transport on {ITER} divertor performance.
\newblock {\em Nuclear Fusion}, 45(7):608--616, July 2005.

\bibitem{kotov_numerical_2007}
V.~Kotov, D.~Reiter, and A.~S. Kukushkin.
\newblock Numerical study of the {ITER} divertor plasma with the {B2}-{EIRENE}
  code package, 2007.

\bibitem{stotler_neutral_1994}
D.~Stotler and C.~Karney.
\newblock Neutral {Gas} {Transport} {Modeling} with {DEGAS} 2.
\newblock {\em Contributions to Plasma Physics}, 34(2-3):392--397, 1994.

\bibitem{reiter_randschicht-konfiguration_1984}
D.~Reiter.
\newblock {Randschicht-Konfiguration von Tokamaks: Entwicklung und Anwendung
  stochastischer Modelle zur Beschreibung des Neutralgastransports}.
\newblock Technical report, Kernforschungsanlage Jülich, 1984.

\bibitem{bhatnagar_model_nodate}
P.~L. Bhatnagar, E.~P. Gross, and M.~Krook.
\newblock A {Model} for {Collision} {Processes} in {Gases}. {I}. {Small}
  {Amplitude} {Processes} in {Charged} and {Neutral} {One}-{Component}
  {Systems}.
\newblock page~15.

\bibitem{krstic_elastic_1998}
P.~S. Krstic and D.~R. Schultz.
\newblock Elastic and {Related} {Transport} {Cross} {Sections} for {Collisions}
  among {Isotopomers} of {H}+ + {H}, {H}+ + {H2}, {H}+ + {He}, {H} + {H}, and
  {H} + {H2}.
\newblock volume~8. International Atomic Energy Agency, Vienna, 1998.

\bibitem{janev_elementary_1988}
R.~Janev, W.~Langer, K.~Evans, and D.~Post.
\newblock Elementary {Processes} in {Hydrogen}-{Helium} {Plasmas}.
\newblock {\em Nuclear Fusion}, 28(3):529, 1988.

\bibitem{bray_electron_2012}
I~Bray.
\newblock Electron- and photon-impact atomic ionisation.
\newblock {\em Physics Reports}, page~40, 2012.

\bibitem{bernard_kinetic_2022}
T.~N. Bernard, F.~D. Halpern, G.~W. Hammett, M.~Francisquez, N.~R. Mandell,
  J.~Juno, G.~J. Wilkie, and J.~Guterl.
\newblock Kinetic modeling of neutral transport for a continuum gyrokinetic
  code.
\newblock {\em Physics of Plasmas}, 29:052501, 2022.

\bibitem{spanier_two_1966}
J.~Spanier.
\newblock Two pairs of families of estimators for transport problems.
\newblock {\em SIAM Journal on Applied Mathematics}, 14(4):702, 1966.

\bibitem{hakim2020continuum}
A.~H. Hakim, N.~R. Mandell, T.~N. Bernard, M.~Francisquez, G.~W. Hammett, and
  E.~L. Shi.
\newblock Continuum electromagnetic gyrokinetic simulations of turbulence in
  the tokamak scrape-off layer and laboratory devices.
\newblock {\em Physics of Plasmas}, 27(4):042304, 2020.

\bibitem{mandell_magnetic_2020}
N.~R. Mandell.
\newblock {\em Magnetic Fluctuations in Gyrokinetic Simulations of Tokamak
  Scrape-Off Layer Turbulence}.
\newblock {PhD} thesis, Princeton University, 2021.

\bibitem{hagelaar_solving_2005}
G.~J.~M. Hagelaar and L.~C. Pitchford.
\newblock Solving the {Boltzmann} equation to obtain electron transport
  coefficients and rate coefficients for fluid models.
\newblock {\em Plasma Sources Science and Technology}, 14(4):722--733, November
  2005.

\bibitem{hagelaar_bolsig_nodate}
G.~J.~M. Hagelaar.
\newblock {BOLSIG}+.
\newblock \url{lxcat.net}.

\bibitem{alves_ist-lisbon_2014}
L.~L. Alves.
\newblock The {IST}-{LISBON} database on {LXCat}.
\newblock In {\em Journal of {Physics}: {Conference} {Series}}, volume 565,
  page 012007. IOP Publishing, 2014.
\newblock Issue: 1.

\bibitem{pitchford_lxcat_2017}
L.~C. Pitchford, L.~L. Alves, K.~Bartschat, S.~F. Biagi, M.-C. Bordage,
  I.~Bray, C.~E. Brion, M.~J. Brunger, L.~Campbell, A.~Chachereau, and
  {others}.
\newblock Lxcat: {An} open-access, web-based platform for data needed for
  modeling low temperature plasmas.
\newblock {\em Plasma Processes and Polymers}, 14(1-2):1600098, 2017.
\newblock Publisher: Wiley Online Library.

\bibitem{kitzler_high_2015}
G.~Kitzler and J.~Sch{\" o}berl.
\newblock {A high order space–momentum discontinuous Galerkin method for the
  Boltzmann equation}.
\newblock {\em Comput. Math. Appl.}, 70(7):1539--1554, 2015.

\bibitem{wilkie_microturbulent_2015}
G.~J. Wilkie.
\newblock {\em Microturbulent transport of non-{Maxwellian} alpha particles}.
\newblock {PhD} thesis, University of Maryland, 2015.

\bibitem{druyvesteyn_influence_1930}
M.~J. Druyvesteyn.
\newblock Influence of energy loss by elastic collisions in the theory of
  electron diffusion.
\newblock {\em Physica}, 10:61, 1930.

\bibitem{morse_velocity_1935}
P.~M. Morse, W.~P. Allis, and E.~S. Lamar.
\newblock Velocity distributions for elastically colliding electrons.
\newblock {\em Physical Review}, 48(5):412, 1935.

\end{thebibliography}

\end{document}